\documentclass[letterpaper]{article} 
\usepackage{aaai23}  
\usepackage{times}  
\usepackage{helvet}  
\usepackage{courier}  
\usepackage[hyphens]{url}  
\usepackage{graphicx} 
\usepackage{natbib}  
\usepackage{caption} 
\usepackage{algorithm}
\usepackage{algorithmic}
\usepackage[table,xcdraw]{xcolor}
\usepackage{multicol}
\usepackage{multirow}
\usepackage[normalem]{ulem}
\useunder{\uline}{\ul}{}
\usepackage{graphicx}
\usepackage{subfigure}
\usepackage{setspace}
\usepackage{booktabs}
\usepackage{makecell}
\usepackage{balance}

\usepackage{newfloat}
\usepackage{listings}
\DeclareCaptionStyle{ruled}{labelfont=normalfont,labelsep=colon,strut=off} 
\floatstyle{ruled}
\newfloat{listing}{tb}{lst}{}
\floatname{listing}{Listing}
\title{FinalMLP: An Enhanced Two-Stream MLP Model for CTR Prediction}
\author{
    Kelong Mao\textsuperscript{\rm 1*}, 
    Jieming Zhu\textsuperscript{\rm 2}\thanks{Both authors contributed equally. Jieming Zhu is the corresponding author.},
    Liangcai Su\textsuperscript{\rm 3},
    Guohao Cai\textsuperscript{\rm 2},
    Yuru Li\textsuperscript{\rm 2},
    Zhenhua Dong\textsuperscript{\rm 2}
}
\affiliations{
    \textsuperscript{\rm 1}Gaoling School of Artificial Intelligence, Renmin University of China\\
    \textsuperscript{\rm 2}Huawei Noah's Ark Lab\\
    \textsuperscript{\rm 3}Tsinghua University\\

    \texttt{kyriemkl@gmail.com, jiemingzhu@ieee.org, sulc21@mails.tsinghua.edu.cn}
}

\begin{document}

\maketitle

\begin{abstract}
 Click-through rate (CTR) prediction is one of the fundamental tasks for online advertising and recommendation. While multi-layer perceptron (MLP) serves as a core component in many deep CTR prediction models, it has been widely recognized that applying a vanilla MLP network alone is inefficient in learning multiplicative feature interactions. As such, many two-stream interaction models (e.g., DeepFM and DCN) have been proposed by integrating an MLP network with another dedicated network for enhanced CTR prediction. As the MLP stream learns feature interactions implicitly, existing research focuses mainly on enhancing explicit feature interactions in the complementary stream. In contrast, our empirical study shows that a well-tuned two-stream MLP model that simply combines two MLPs can even achieve surprisingly good performance, which has never been reported before by existing work. Based on this observation, we further propose feature gating and interaction aggregation layers that can be easily plugged to make an enhanced two-stream MLP model, FinalMLP. In this way, it not only enables differentiated feature inputs but also effectively fuses stream-level interactions across two streams. Our evaluation results on four open benchmark datasets as well as an online A/B test in our industrial system show that FinalMLP achieves better performance than many sophisticated two-stream CTR models. Our source code will be available at \textcolor{magenta}{\url{https://reczoo.github.io/FinalMLP}}.


\end{abstract}

\section{Introduction}
\label{sec:intro}
Click-through rate (CTR) prediction is a fundamental task in online advertising and recommender systems~\cite{WideDeep,GBDT_LR}. 
The accuracy of CTR prediction not only has a direct effect on user engagement but also significantly influences the revenue of business providers. 
One of the key challenges in CTR prediction is to learn complex relationships among features such that a model can still generalize well in case of rare feature interactions. Multi-layer perceptron (MLP), as a powerful and versatile component in deep learning, has become a core building block of various CTR prediction models~\cite{fuxictr}. Although MLP is known to be a universal approximator in theory, it has been widely recognized that in practice applying a vanilla MLP network is inefficient to learn multiplicative feature interactions (e.g., dot)~\cite{DCN,DCN_V2,NCFvsMF}. 

To enhance the capability of learning explicit feature interactions (2nd- or 3rd-order features), a variety of feature interaction networks have been proposed. Typical examples include factorization machines (FM)~\cite{FM}, cross network~\cite{DCN}, compressed interaction network (CIN)~\cite{xDeepFM}, self-attention based interaction~\cite{autoint}, adaptive factorization network (AFN)~\cite{AFN}, and so on. These networks introduce inductive bias for learning feature interactions efficiently but lose the expressiveness of MLP as our experiments shown in Table~\ref{tab::single_model_comparison}. As such, two-stream CTR prediction models have been widely employed, such as DeepFM~\cite{DeepFM}, DCN~\cite{DCN}, xDeepFM~\cite{xDeepFM}, and AutoInt+~\cite{autoint}, which integrate both an MLP network and a dedicated feature interaction network together for enhanced CTR prediction. Concretely, the MLP stream learns feature interactions implicitly, while the other stream enhances explicit feature interactions in a complementary way. Due to their effectiveness, two-stream models have become a popular choice for industrial deployment~\cite{CTR_Survey}.

Although many existing studies have validated the effectiveness of two-stream models against a single MLP model, none of them reports a performance comparison to a two-stream MLP model that simply combines two MLP networks in parallel (denoted as DualMLP). Therefore, our work makes the first effort to characterize the performance of DualMLP. Our empirical study on open benchmark datasets shows that DualMLP, despite its simplicity, can achieve surprisingly good performance, which is comparable to or even better than many well-designed two-stream models (see our experiments). This observation motivates us to study the potential of such a two-stream MLP model and further extend it to build a simple yet strong model for CTR prediction.

Two-stream models in fact can be viewed as an ensemble of two parallel networks. One advantage of these two-stream models is that each stream can learn feature interactions from a different perspective and thus complements each other to achieve better performance. For instance, Wide\&Deep~\cite{WideDeep} and
DeepFM~\cite{DeepFM} propose to use one stream to capture low-order feature interactions and another to learn high-order feature interactions. DCN~\cite{DCN} and AutoInt+~\cite{autoint} advocate learning explicit feature interactions and implicit feature interactions in two streams respectively. xDeepFM~\cite{xDeepFM} further enhances feature interaction learning from vector-wise and bit-wise perspectives. These previous results verify that the differentiation (or diversity) of two network streams makes a big impact on the effectiveness of two-stream models.

Compared to the existing two-stream models that resort to designing different network structures (e.g., CrossNet~\cite{DCN} and CIN~\cite{xDeepFM}) to enable stream differentiation, DualMLP is limited in that both streams are simply MLP networks. Our preliminary experiments also reveal that DualMLP, when tuned with different network sizes (w.r.t., number of layers or units) for two MLPs, can achieve better performance. This result promotes us to further explore how to enlarge the differentiation of two streams to improve DualMLP as a base model. In addition, existing two-stream models often combine two streams via summation or concatenation, which may waste the opportunity to model the high-level (i.e., stream-level) feature interactions. How to better fuse the stream outputs becomes another research problem that deserves further exploration.



To address these problems, in this paper, we build an enhanced two-stream MLP model, namely FinalMLP, which integrates \underline{\textbf{f}}eature gating and \underline{\textbf{in}}teraction \underline{\textbf{a}}ggregation \underline{\textbf{l}}ayers on top of two \underline{\textbf{MLP}} module networks. More specifically, we propose a stream-specific feature gating layer that allows obtaining gating-based feature importance weights for soft feature selection. That is, the feature gating can be computed from different views via conditioning on learnable parameters, user features, or item features, which produces global, user-specific, or item-specific feature importance weights respectively. By flexibly choosing different gating-condition features, we are able to derive stream-specific features for each stream and thus enhance the differentiation of feature inputs for complementary feature interaction learning of two streams. To fuse the stream outputs with stream-level feature interaction, we propose an interaction aggregation layer based on second-order bilinear fusion~\cite{BilinearCNN,BilinearModel}. To reduce the computational complexity, we further decompose the computation into $k$ sub-groups, which leads to efficient multi-head bilinear fusion. 
Both feature gating and interaction aggregation layers can be easily plugged into existing two-stream models.

Our experimental results on four open benchmark datasets show that FinalMLP outperforms the existing two-stream models and attains new state-of-the-art performance. Furthermore, we validate its effectiveness in industrial settings through both offline evaluation and online A/B testing, where FinalMLP also shows significant performance improvement over the deployed baseline. We envision that the simple yet effective FinalMLP model could serve as a new strong baseline for future developments of two-stream CTR models. 
The main contributions of this paper are summarized as follows:



\begin{itemize}
    \item To our knowledge, this is the first work that empirically demonstrates the surprising effectiveness of a two-stream MLP model, which may be contrary to popular belief in the literature.
    \item We propose FinalMLP, an enhanced two-stream MLP model with pluggable feature gating and interaction aggregation layers.
    \item Both Offline experiments on benchmark datasets and an online A/B test in production systems have been conducted to validate the effectiveness of FinalMLP.
\end{itemize}
\section{Background and Related Work}

\begin{figure*}[!t]
	\centering
	\includegraphics[width=0.9\textwidth]{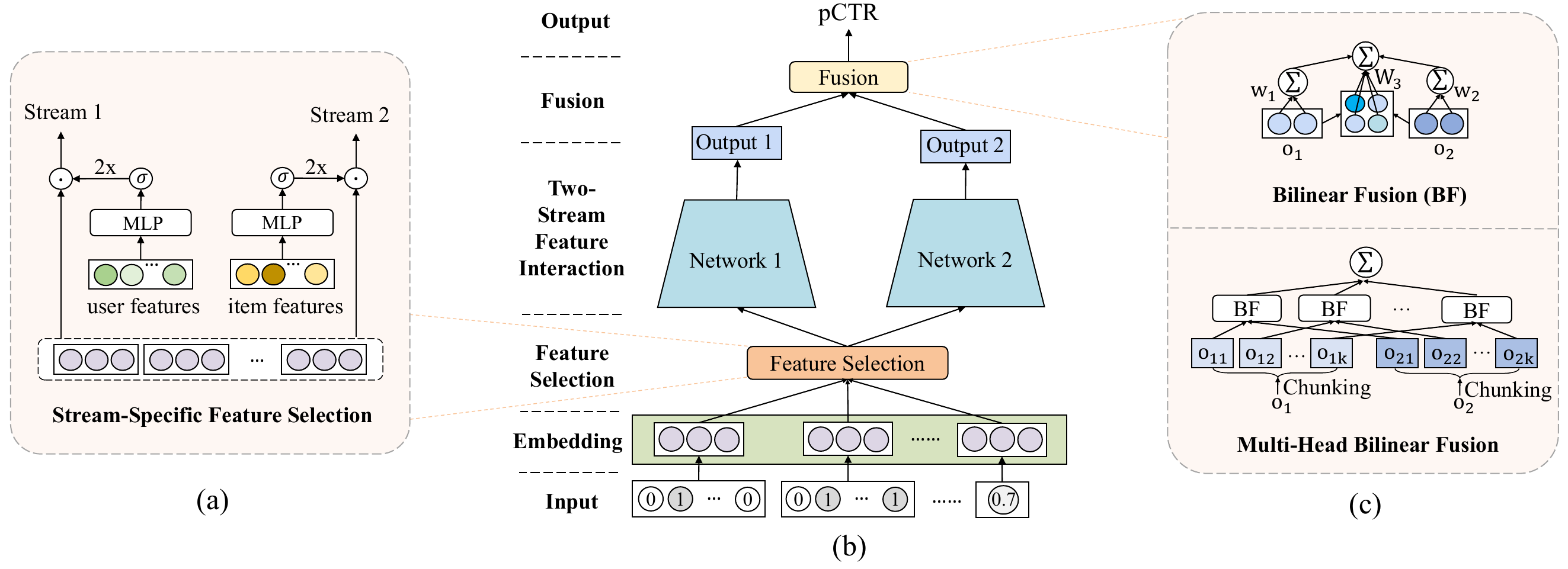}
	\caption{(a) An illustration of stream-specific feature selection. (b) A general framework of two-stream CTR models. (c) The multi-head bilinear fusion.}
	\label{fig:two_stream_model_overview}
\end{figure*}

In this section, we briefly review the framework and representative two-stream models for CTR prediction.

\subsection{Framework of Two-Stream CTR Models}
\label{sec:framework}
We illustrate a framework of two-stream CTR models in Figure~\ref{fig:two_stream_model_overview}(b), which consists of the following key components.

\subsubsection{\textbf{Feature Embedding}}
Embedding is a common way to map high-dimensional and sparse raw features into dense numeric representations. Specifically, suppose that the raw input feature is $x = \{x_1, ..., x_M\}$ with $M$ feature fields, where $x_i$ is the feature of the $i$-th field. In general, $x_i$ can be a \textit{categorical}, \textit{multi-valued}, or \textit{numerical} feature. Each of them can be transformed into embedding vectors accordingly. Interested readers may refer to~\cite{fuxictr} for more details on feature embedding methods. Then, these feature embeddings will be concatenated and fed into the following layer.





\subsubsection{\textbf{Feature Selection}}
Feature selection is an \textit{optional} layer in the framework of two-stream CTR models. In practice, feature selection is usually performed through offline statistical analysis or model training with difference comparison~\cite{FeatureSelection}. Instead of hard feature selection, in this work, we focus on the soft feature selection through the feature gating mechanism~\cite{FiBiNET,FeatureGating}, which aims to obtain feature importance weights to help amplify important features while suppressing noisy features. In this work, we study stream-specific feature gating to enable differentiated stream inputs.

\subsubsection{\textbf{Two-Stream Feature Interaction}}
\label{sec:fusion_layer}
The key feature of two-stream CTR models is to employ two parallel networks to learn feature interactions from different views. Basically, each stream can adopt any type of feature interaction network (e.g., FM~\cite{FM}, CrossNet~\cite{DCN}, and MLP). Existing work typically applies two different network structures to the two streams in order to learn complementary feature interactions (e.g., explicit vs. implicit, bit-wise vs. vector-wise). In this work, we make the first attempt to use two MLP networks as two streams.


\subsubsection{\textbf{Stream-Level Fusion}}
Stream-level fusion is required to fuse the outputs of two streams to obtain the final predicted click probability $\hat{y}$. 
Suppose $\mathbf{o_1}$ and $\mathbf{o_2}$ as two output representations, it can be formulated as:
$\hat{y} = \sigma (w^T\mathcal{F}(\mathbf{o_1}, \mathbf{o_2}))$,
where $\mathcal{F}$ denotes the fusion operation which is commonly set as summation or concatenation\footnote{In case that the output dimension is $1$, the fusion operation of concatenation is approximately equivalent to summation because $w^T[o_1,o_2] = [w_1,w_2]^T[o_1,o_2] = w_1^To_1 + w_2^To_2$.}. $w$ denotes a linear function to map the output dimension to $1$ when necessary. $\sigma$ is the sigmoid function. Existing work only performs a first-order linear combination of stream outputs, so it fails to mine the stream-level feature interactions. In this work, we explore a second-order bilinear function fo stream-level interaction aggregation.


\subsection{Representative Two-Stream CTR Models} \label{sec:review_of_two_stream_model}
We summarize some representative two-stream models that cover a wide spectrum of studies on CTR prediction. 
\begin{itemize}
    \item \textbf{Wide\&Deep}: Wide\&Deep~\cite{WideDeep} is a classical two-stream feature interaction learning framework that combines a generalized linear model (wide stream) and an MLP network (deep stream).
    \item \textbf{DeepFM}: DeepFM~\cite{DeepFM} extends Wide\&Deep by replacing the wide stream with FM to learn second-order feature interactions explicitly.
    \item \textbf{DCN}: In DCN~\cite{DCN}, a cross network is proposed as one stream to perform  high-order feature interactions in an explicit way, while another MLP stream learns feature interactions implicitly. 
    \item \textbf{xDeepFM}: xDeepFM~\cite{xDeepFM} employs a compressed interaction network (CIN) to capture high-order feature interactions in a vector-wise way, and also adopts MLP as another stream to learn bit-wise feature interactions. 
    \item \textbf{AutoInt+}: AutoInt~\cite{autoint} applies self-attention networks to learning high-order feature interactions. AutoInt+ integrates AutoInt and MLP as two complementary streams.
    \item \textbf{AFN+}: AFN~\cite{AFN} leverages logarithmic transformation layers to learn adaptive-order feature interactions. AFN+ combines AFN with MLP in a two-stream manner.
    \item \textbf{DeepIM}: In DeepIM~\cite{DeepIM}, an interaction machine (IM) module is proposed to efficiently compute high-order feature interactions. It uses IM and MLP separately in two streams.
    \item \textbf{MaskNet}: In MaskNet~\cite{MaskNet}, a MaskBlock is proposed by combining layer normalization, instance-guided mask, and feed-forward layer. The parallel MaskNet is a two-stream model that uses two MaskBlocks in parallel.
    \item \textbf{DCN-V2}: DCN-V2~\cite{DCN_V2} improves DCN with a more expressive cross network to better capture explicit feature interactions. It still uses MLP as another stream in the parallel version.
    \item \textbf{EDCN}: EDCN~\cite{EDCN} is not a strict two-stream model, since it proposes a bridge module and a regulation module to bridge the information fusion between hidden layers of two streams. However, its operations limit each stream to having the same size of hidden layers and units, reducing flexibility.
\end{itemize}

\section{Our Model}

In this section, we first present the simple two-stream MLP base model, \textbf{DualMLP}. Then, we describe two pluggable modules, feature gating and interaction aggregation layers, which results in our enhanced model, \textbf{FinalMLP}.


\subsection{Two-Stream MLP Model}
Despite its simplicity, to the best of our knowledge, the two-stream MLP model  has not been reported before by previous work. Thus, we make the first effort to study such a model for CTR prediction, denoted as DualMLP, which simply combines two independent MLP networks as two streams. Specifically, the two-stream MLP model can be formulated as follows:
\begin{eqnarray}
\mathbf{o_1} &=& {MLP}_1(\mathbf{h_1}), \\
\mathbf{o_2} &=& {MLP}_2(\mathbf{h_2}),
\end{eqnarray}
where ${MLP}_1$ and ${MLP}_2$ are two MLP networks. The size of both MLPs (w.r.t. hidden layers and units) can be set differently according to data. $\mathbf{h_1}$ and $\mathbf{h_2}$ denote the feature inputs while $\mathbf{o_1}$ and $\mathbf{o_2}$ are the output representations of the two streams, respectively. 

In most previous work~\cite{DCN,DeepFM,AFN}, the feature inputs $\mathbf{h_1}$ and $\mathbf{h_2}$ are usually set as the same one, which is a concatenation of feature embeddings $\mathbf{e}$ (optionally with some pooling), i.e., $\mathbf{h_1}=\mathbf{h_2}=\mathbf{e}$. Meanwhile, the stream outputs are often fused via simple operations, such as summation and concatenation, ignoring stream-level interactions. In the following, we introduce two modules that can be plugged into the inputs and outputs respectively to enhance the two-stream MLP model.

\subsection{Stream-Specific Feature Selection}
\label{sec:input_differentiation}
Many existing studies~\cite{DeepFM,xDeepFM,DCN,autoint}
highlight the effectiveness of combining two different feature interaction networks (e.g., implicit vs. explicit, low-order vs. high-order, bit-wise vs. vector-wise) to achieve accurate CTR prediction. Instead of designing specialized network structures, our work aims to enlarge the difference between two streams through stream-specific feature selection, which produces differentiated feature inputs. 

Inspired by the gating mechanism used in MMOE~\cite{MMoE}, we propose a stream-specific feature gating module to soft-select stream-specific features, i.e., re-weighting feature inputs differently for each stream. In MMOE, gating weights are conditioned on task-specific features to re-weight expert outputs. Likewise, we perform feature gating from different views via conditioning on learnable parameters, user features, or item features, which produces global, user-specific, or item-specific feature importance weights respectively. 


Specifically, we make stream-specific feature selection through the context-aware feature gating layer as follows.
\begin{eqnarray}
&&\mathbf{g_1} = {Gate}_{1}(\mathbf{x_1}),~~~\mathbf{g_2} = {Gate}_{2}(\mathbf{x_2}), \\
&&\mathbf{h_1} = 2\sigma (\mathbf{g_1})\odot \mathbf{e},~~\mathbf{h_2} = 2\sigma(\mathbf{g_2})\odot\mathbf{e},
\end{eqnarray}
where ${Gate}_{i}$ denotes an MLP-based gating network, which takes stream-specific conditional features $\mathbf{x_i}$ as input and outputs element-wise gating weights $\mathbf{g_i}$. Note that it is flexible to either choose $\mathbf{x_i}$ from a set of user/item features or set it as learnable parameters. The feature importance weights are obtained by using the sigmoid function $\sigma$ and a multiplier of $2$ to transform them to the range of $[0,2]$ with an average of $1$. Given the concatenated feature embeddings $\mathbf{e}$, we can then obtain weighted feature outputs $\mathbf{h_1}$ and $\mathbf{h_2}$ via element-wise product $\odot$. 

Our feature gating module allows making differentiated feature inputs for two streams by setting conditional features $\mathbf{x_i}$ from a different view. For example, Figure~\ref{fig:two_stream_model_overview}(a) demonstrates a case of user- and item-specific feature gating, which modulates each stream from the view of users and items, respectively. This reduces ``homogeneous'' learning between two similar MLP streams, and would enable more complementary learning of feature interactions.

\subsection{Stream-Level Interaction Aggregation}
\subsubsection{Bilinear Fusion} As mentioned before, existing work mostly employs summation or concatenation as the fusion layer, but these operations fail to capture stream-level feature interactions. Inspired by the widely studied bilinear pooling in the CV domain~\cite{BilinearCNN,BilinearModel}, we propose a bilinear interaction aggregation layer to fuse the stream outputs
with stream-level feature interaction. As illustrated in Figure~\ref{fig:two_stream_model_overview}(c), the predicted click probability is formulated as follows.
\begin{eqnarray}
\hat{y} = \sigma(b + \mathbf{w}_1^{T}\mathbf{o}_1 + \mathbf{w}_2^{T}\mathbf{o}_2 + \mathbf{o}_1^{T}\mathbf{W}_3\mathbf{o}_2),
\end{eqnarray}\label{equ:bilinear}
where $b \in \mathcal{R}$, $\mathbf{w}_1\in \mathcal{R}^{d_1 \times 1}$, $\mathbf{w}_2 \in \mathcal{R}^{d_2 \times 1}$, $\mathbf{W}_3 \in \mathcal{R}^{d_1 \times d_2}$ are learnable weights. $d_1$ and $d_2$ denote the dimensions of $\mathbf{o}_1$ and $\mathbf{o}_2$, respectively.
The bilinear term $\mathbf{o}_1^{T}\mathbf{W_3}\mathbf{o}_2$ models the second-order interactions between $\mathbf{o}_1$ and $\mathbf{o}_2$. Especially, when $\mathbf{W}_3$ is an identity matrix, the term simulates the dot product. When setting $\mathbf{W}_3$ to a zero matrix, it degenerates to the traditional concatenation fusion with a linear layer, i.e., $b + [\mathbf{w}_1,\mathbf{w}_2]^T[\mathbf{o}_1,\mathbf{o}_2]$. 

Interestingly, the bilinear fusion also has a connection to the commonly used FM model. Concretely, FM models the second-order feature interactions among a $m$-dimensional input feature vector $\mathbf{x}$ (via one-hot/multi-hot feature encoding and concatenation) for CTR prediction by:
\begin{eqnarray}
\hat{y} = \sigma(b + \mathbf{w}^{\top}\mathbf{x} + \mathbf{x}^{\top} \texttt{upper}\big(\mathbf{PP^{\top}})\mathbf{x}\big),
\end{eqnarray}
where $b \in \mathcal{R}$, $\mathbf{w} \in \mathcal{R}^{m \times 1}$, $\mathbf{P} \in \mathcal{R}^{m \times d}$ are learnable weights with $d \ll m$, and \texttt{upper} selects the strictly upper triangular part of the matrix~\cite{FM}. As we can see, FM is a special form of bilinear fusion when $\mathbf{o}_1=\mathbf{o}_2$.

However, when $\mathbf{o}_1$ and $\mathbf{o}_2$ are high-dimensional, it is parameter-intensive and computation-expensive to compute Equation (5). For example, to fuse two 1000-dimensional outputs, $\mathbf{W}_3 \in \mathcal{R}^{1000 \times 1000}$ takes up 1 million parameters and its computation becomes expensive. To reduce the computational complexity, we introduce our extended multi-head bilinear fusion in the following.


\subsubsection{Multi-Head Bilinear Fusion} 
Multi-head attention is appealing for the ability to combine knowledge of the same attention pooling from different representation subspaces. It leads to reduced computation and consistent performance improvements in the recent successful transformer models~\cite{Transformer}. Inspired by its success, we extend the bilinear fusion to a multi-head version. Specifically, instead of directly computing the bilinear term in Equation (5), we chunk each of $\mathbf{o_1}$ and $\mathbf{o_2}$ into $k$ subspaces:
\begin{eqnarray}
\mathbf{o_1} &=& [\mathbf{o_{11}}, ..., \mathbf{o_{1k}} ], \\
\mathbf{o_2} &=& [\mathbf{o_{21}}, ..., \mathbf{o_{2k}} ],
\end{eqnarray}
where $k$ is a hyper-parameter and $\mathbf{o_{ij}}$ denotes the $j$-th subspace representation of the $i$-th output vector ($i \in \{1, 2\}$). Similar to multi-head attention, we perform the bilinear fusion in each subspace that pairs $\mathbf{o_{1j}}$ and $\mathbf{o_{2j}}$ as a group. Then, we aggregate the subspace computation by sum pooling to get the final predicted click probability:
\begin{eqnarray}
\hat{y} &=& \sigma(\sum_{j=1}^{k}{BF}(\mathbf{o_{1j}}, \mathbf{o_{2j}})),
\end{eqnarray}
where ${BF}$ denotes the bilinear fusion without sigmoid activation in Equation (5).


Through the subspace computation as with multi-head attention, we can theoretically reduce the number of parameters and the computation complexity of bilinear fusion by a factor of $k$, i.e., from $\mathcal{O}(d_1d_2)$ to $\mathcal{O}(\frac{d_1d_2}{k})$. Especially, when setting $k=d_1=d_2$, it degenerates to an element-wise product fusion.
If $k=1$, it equals to the original bilinear fusion. Selecting an appropriate $k$ realizes multi-head learning so that the model may achieve better performance. In practice, the multi-head fusions for $k$ subspaces are computed in parallel in GPUs, which further increases efficiency.


Finally, our stream-specific feature gating and stream-level interaction aggregation modules can be plugged to produce an enhanced two-stream MLP model, FinalMLP.

\subsection{Model Tranining}
To train FinalMLP, we apply the widely used binary cross-entropy loss:
$\mathcal{L} = - \frac{1}{N} \sum\big (y {log}\hat{y} + (1-y) {log}(1 - \hat{y})\big )$,
where $y$ and $\hat{y}$ denote the true label and the estimated click probability respectively from each of a total of $N$ samples. 
\section{Experiments}\label{sec:exp}

\subsection{Experimental Setup}

\subsubsection{Datasets}
We experiment with four open benchmark datasets, including Criteo, Avazu, MovieLens, and Frappe. We reuse the preprocessed data by~\cite{AFN} and follow the same settings on data splitting and preprocessing. Table~\ref{tab:dataset} summairies the statistics of the datasets. 



\begin{table}[!t]
\centering
\caption{The statistics of open datasets.}
\scalebox{0.9}{\begin{tabular}{c|c|c|c}
\toprule
Dataset   & \#Instances & \#Fields & \#Features \\ \midrule
Criteo    & 45,840,617  & 39       & 2,086,936  \\
Avazu     & 40,428,967  & 22       & 1,544,250  \\
MovieLens & 2,006,859   & 3        & 90,445     \\
Frappe    & 288,609     & 10       & 5,382      \\ \bottomrule
\end{tabular}}
\label{tab:dataset}
\end{table}

\subsubsection{Evaluation Metric} 
We employ AUC as one of the most widely used evaluation metrics for CTR prediction. Moreover, a 0.1-point increase in AUC is recognized as a significant improvement in CTR prediction~\cite{WideDeep, AFN, DCN_V2}.

\subsubsection{Baselines} 
First, we study a set of single-stream explicit feature interaction networks as follows.
\begin{itemize}
    \item First-order: Logistic Regression (LR)~\cite{LR}.
    \item Second-order: FM~\cite{FM}, AFM~\cite{AFM}, FFM~\cite{FFM}, FwFM~\cite{FwFM}, and FmFM~\cite{FmFM}.
    \item Third-order: HOFM (3rd)~\cite{HighFM}, CrossNet (2L)~\cite{DCN}, CrossNetV2 (2L)~\cite{DCN_V2}, and CIN (2L)~\cite{xDeepFM}. We specially set the maximal order to ``3rd" or the number of interaction layers to ``2L" to obtain third-order feature interactions.
    \item Higher-order: CrossNet~\cite{DCN}, CrossNetV2~\cite{DCN_V2}, CIN~\cite{xDeepFM}, AutoInt~\cite{autoint}, FiGNN~\cite{FiGNN}, AFN~\cite{AFN}, and SAM~\cite{SAM}, which automatically learn high-order feature interactions. 
\end{itemize}

Then, we study a set of representative two-stream CTR models as introduced in the related work section.

\subsubsection{Implementation}
We reuse the baseline models and implement our models based on FuxiCTR~\cite{fuxictr}, an open-source CTR prediction library\footnote{\url{https://reczoo.github.io/FuxiCTR}}. Our evaluation follows the same experimental settings with AFN~\cite{AFN}, by setting the embedding dimension to $10$, batch size to $4096$, and the default MLP size to $[400, 400, 400]$. For DualMLP and FinalMLP, we tune the two MLPs in 1$\sim$3 layers to enhance stream diversity. We set the learning rate to $1e-3$ or $5e-4$. We tune all the other hyper-parameters (e.g., embedding regularization and dropout rate) of all the studied models via extensive grid search (about 30 runs per model on average). 
We note that through optimized FuxiCTR implementation and sufficient hyper-parameter tuning, we obtain much better model performance than what was reported in~\cite{AFN}\footnote{\url{https://github.com/WeiyuCheng/AFN-AAAI-20/issues/11}}. Thus, we report our own experimental results instead of reusing theirs to make a fair comparison. To promote reproducible research, we open sourced the code and running logs of FinalMLP and all the baselines used.

\begin{table*}[!t]
\renewcommand\arraystretch{1.1}
\setlength{\tabcolsep}{2.3pt}
\centering
\caption{Performance comparison of two-stream models for CTR prediction. The best results are in \textbf{bold} and the second-best results are \underline{underlined}.}
\scalebox{0.89}{\begin{tabular}{c|c|cccccccccc|cc}
\toprule
Dataset & Metric & WideDeep & DeepFM & DCN & xDeepFM & AutoInt+ & AFN+ & DeepIM & MaskNet & DCNv2 & EDCN & DualMLP & FinalMLP \\ \midrule
\multirow{2}{*}{Criteo} & AUC &  81.38 & 81.38 & 81.39 & 81.39 & 81.39 & { 81.43} & 81.40 & 81.39 & 81.42 & \underline{81.47} & 81.42 & \textbf{81.49} \\
 & Std & 5.7e-5 & 8.0e-5 & 4.9e-5  & 9.5e-5 & 1.4e-4 &5.9e-5  & 5.9e-5 & 1.3e-4 & 2.0e-4 &  6.6e-5 & 5.6e-4 & 1.7e-4   \\ \hline
\multirow{2}{*}{Avazu} & AUC & 76.46 & 76.48 & 76.47 & 76.49 & 76.45 & 76.48 & {76.52} & 76.49 & { 76.54}  & 76.52 & \underline{76.57} & \textbf{76.66} \\
 & Std &5.4e-4  & 4.4e-4 & 1.2e-3 & 4.1e-4 & 5.2e-4 & 3.7e-4 & 9.2e-5 & 2.6e-3 & 4.7e-4 & 3.0e-4  & 3.5e-4 & 4.9e-4   \\ \hline
\multirow{2}{*}{MovieLens} & AUC & 96.80 & 96.85 & 96.87 & { 96.97} & 96.92 & 96.42 & 96.93 & 96.87 & 96.91  & 96.71 & \underline{96.98} & \textbf{97.20} \\
 & Std & 3.2e-4  & \multicolumn{1}{c}{1.6e-4} & \multicolumn{1}{c}{5.5e-4} & 9.0e-4 & \multicolumn{1}{c}{4.4e-4} & \multicolumn{1}{c}{5.8e-4} & \multicolumn{1}{c}{5.8e-4} & 2.8e-4 & \multicolumn{1}{c}{3.6e-4} & \multicolumn{1}{c|}{3.4e-4} & \multicolumn{1}{c}{4.3e-4} &
 \multicolumn{1}{c}{1.8e-4}  \\ \hline
\multirow{2}{*}{Frappe} & AUC &  98.41 & 98.42 & 98.39 & 98.45 & { 98.48} & 98.26 & 98.44 & 98.43 & 98.45  & \underline{98.50} & {98.47} & \textbf{98.61} \\
 & Std &  7.9e-4 & 1.6e-4  & 3.1e-4 & 3.7e-4 & 7.9e-4 & 1.4e-3 &  6.3e-4 &  5.7e-4  & 4.3e-4 & 5.1e-4 & 3.5e-4 & 1.7e-4  \\ \bottomrule
\end{tabular}}
\label{tab:two_stream_model_comparison}
\vspace{-2ex}
\end{table*}

\subsection{MLP vs. Explicit Feature Interactions}
\label{exp:single_model}
\begin{table}[!t]
\renewcommand\arraystretch{1.15}
\setlength{\tabcolsep}{2.5pt}
\centering
\caption{Performance comparisons between MLP and explicit feature interaction networks. The best results w.r.t. AUC are in \textbf{bold} and the second-best results are \underline{underlined}.}
\scalebox{0.9}{\begin{tabular}{c|c|cccc}
\toprule
Class & Model & Criteo & Avazu & MovieLens & Frappe \\ \midrule
\begin{tabular}[c]{@{}c@{}}First-Order\end{tabular} & LR & 78.86 & 75.16 & 93.42 & 93.56 \\ \hline
 & FM & 80.22 & 76.13 & 94.34 & 96.71 \\
 & AFM & 80.44 & 75.74 & 94.72 & 96.97 \\
 & FFM & 80.60 & {76.25} & 95.22 & 97.88 \\
 & FwFM & 80.63 & 76.02 & 95.58 & 97.76 \\
\multirow{-5}{*}{\begin{tabular}[c]{@{}c@{}}Second-\\ Order\end{tabular}} & FmFM & 80.56 & 75.95 & 94.65 & 97.49 \\ \hline
 & HOFM(3rd) & 80.55 & 76.01 & 94.55 & 97.42 \\
 & CrossNet(2L) & 79.47 & 75.45 & 93.85 & 94.19 \\
 & CrossNetV2(2L) & 81.10 & 76.05 & 95.83 & 97.16 \\
\multirow{-4}{*}{\begin{tabular}[c]{@{}c@{}}Third-\\ Order\end{tabular}} & CIN(2L) & 80.96 &  76.26 & 96.02 & 97.76 \\ \hline
 & CrossNet & 80.41 & 75.97 & 94.40 & 95.94 \\
 & CrossNetV2 & 81.27 & {76.25} & 96.06 & 97.29 \\
 & CIN & 81.17 & 76.24 &  \underline{96.74} & 97.82 \\
 & AutoInt & 81.26 & 76.24 & 96.63 &  \underline{98.31} \\
 & FiGNN & {\underline{81.34}} & 76.22 & 95.25 & 97.61 \\
 & AFN & 81.07 & 75.47 & 96.11 & 98.11 \\ 
  & SAM & {81.31} & \textbf{76.32} & 96.31 & 98.01 \\ \cline{2-6} 
\multirow{-7}{*}{\begin{tabular}[c]{@{}c@{}}Higher-\\ Order\end{tabular}} & {MLP} & {\textbf{81.37}} & \underline{76.30} & \textbf {96.78} & \textbf{98.33} \\ \bottomrule
\end{tabular}}
\label{tab::single_model_comparison} 
\vspace{-2ex}
\end{table}

While feature interaction networks have been widely studied, there is a lack of comparison between MLP and well-designed feature interaction networks. Previous work proposes many explicit feature interaction networks, e.g., cross network~\cite{DCN}, CIN~\cite{xDeepFM}, AutoInt~\cite{autoint}, and AFN~\cite{AFN}, to overcome the limitation of MLP in learning high-order feature interactions. Yet, most of these studies fail to directly compare explicit feature interaction networks with MLP (a.k.a, DNN or YouTubeDNN~\cite{DNN}) alone, but only evaluate the effectiveness of two-stream model variants (e.g., DCN, xDeepFM, and AutoInt+) against MLP. In this work, we make such a comparison in Table~\ref{tab::single_model_comparison}. 
We enumerate the representative methods that are used for first-order, second-order, third-order, and higher-order feature interactions. Surprisingly, we observe that MLP can perform neck to neck with or even outperform the well-designed explicit feature interaction networks. For example, MLP achieves the best performance on Criteo, MovieLens, and Frappe, while attaining the second-best performance on Avazu, with only an AUC gap of 0.02 points compared to SAM. The observation is also consistent with the results reported in~\cite{DCN_V2}, where a well-tuned MLP model (i.e., DNN) is shown to obtain comparable performance with many existing models. 

Overall, the strong performance achieved by MLP indicates that, despite its simple structure and weakness in learning multiplicative features, MLP is very expressive in learning feature interactions in an implicit way. This also partially explains why existing studies tend to combine both explicit feature interaction networks with MLP as a two-stream model for CTR prediction. Unfortunately, its strength has never been explicitly revealed in any existing work. Inspired by the above observations, we make one step further to study the potential of an unexplored model structure that simply takes two MLPs as a two-stream MLP model.

\subsection{DualMLP and FinalMLP vs. Two-Stream Baselines}
Following existing studies, we make a thorough comparison of representative two-stream models as shown in Table~\ref{tab:two_stream_model_comparison}. From the results, we have the following observations:

First, we can see that two-stream models generally outperform the single-stream baselines reported in Table~\ref{tab::single_model_comparison}, especially the single MLP model. This conforms to existing work, which reveals that two-stream models can learn complementary features and thus enable better modeling for CTR prediction.

Second, the simple two-stream model, DualMLP, performs surprisingly well. With careful tuning of MLP layers of the two streams, DualMLP can achieve comparable or even better performance compared to the other sophisticated two-stream baselines. To the best of our knowledge, the strong performance of DualMLP has never been reported in the literature. In our experiments, we found that increasing stream network diversity by setting different MLP sizes in two streams can improve the performance of DualMLP. This motivates us to further develop an enhanced two-stream MLP model, FinalMLP. 

Third, through our pluggable extensions on feature gating and fusion, FinalMLP consistently outperforms DualMLP as well as all the other compared two-stream baselines across the four open datasets. In particular, FinalMLP significantly surpasses the existing strongest two-stream models by 0.12 points (DCNv2), 0.23 points (xDeepFM), and 0.11 points (AutoInt+) in AUC on Avazu, MovieLens, and Frappe, respectively. This demonstrates the effectiveness of our FinalMLP. As of the time of writing, FinalMLP ranks the 1st on the CTR prediction leaderboards of PapersWithCode\footnote{\url{https://paperswithcode.com/sota/click-through-rate-prediction-on-criteo}} and BARS\footnote{\url{https://openbenchmark.github.io/BARS/CTR}}~\cite{BARS} on Criteo.

\begin{figure*}[!t]
	\centering
		\subfigure[Avazu]{
		\includegraphics[width=0.58\columnwidth]{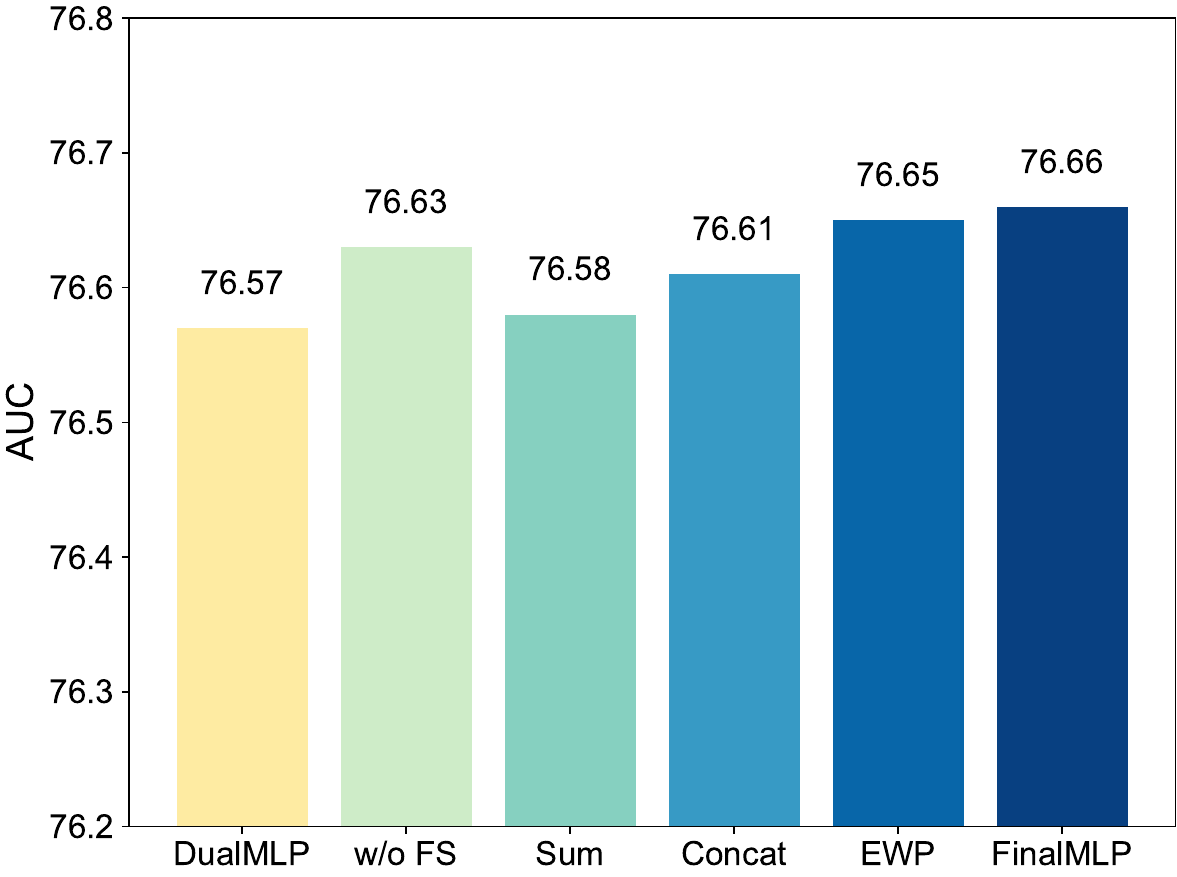}}
    	\subfigure[MovieLens]{
    		\includegraphics[width=0.58\columnwidth]{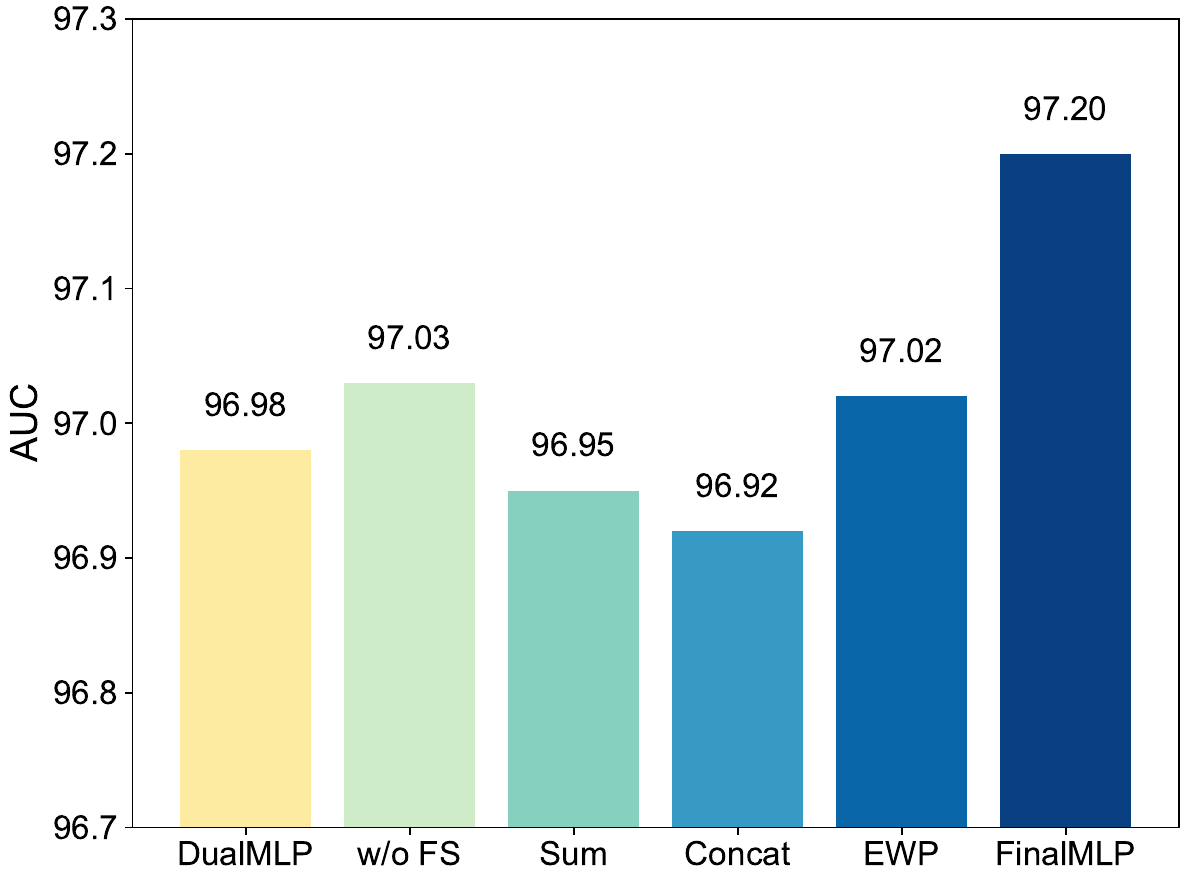}}
		\subfigure[Frappe]{
		\includegraphics[width=0.58\columnwidth]{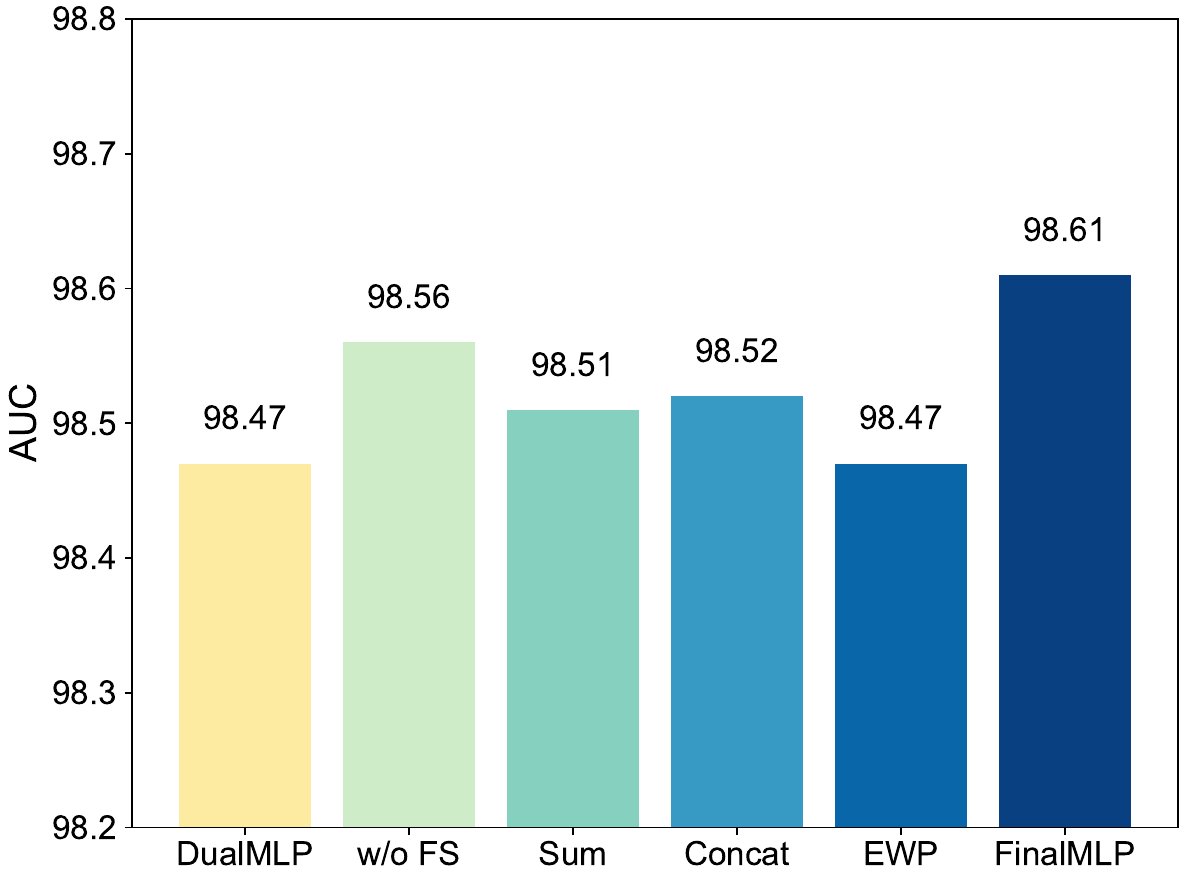}}
  \vspace{-2ex}
	\caption{The ablation study results of FinalMLP.}
	\vspace{-1ex}
 \label{fig:ablation_studies}
\end{figure*}

\subsection{Ablation Studies}
In this section, ablation studies are shown to investigate the effects of the important designs of FinalMLP. 
\subsubsection{Effects of Feature Selection and Bilinear Fusion}
Specifically, we compare FinalMLP with the following variants:
\begin{itemize}
    \item \textbf{DualMLP}: the simple two-stream MLP model that simply takes two MLPs as two streams.
    \item \textbf{w/o FS}: FinalMLP without the stream-specific feature selection module via context-aware feature gating.
    \item \textbf{Sum}: Using the summation fusion in FinalMLP. 
    \item \textbf{Concat}: Using the concatenation fusion in FinalMLP. 
    \item \textbf{EWP}: Using the {E}lement-{W}ise {P}roduct (i.e., Hadamard product) fusion in FinalMLP. 
\end{itemize}
The ablation study results are presented in Figure~\ref{fig:ablation_studies}. 
We can see that the performance drops when removing the feature selection module or replacing the bilinear fusion with other commonly used fusion operations. This verifies the effectiveness of our feature selection and bilinear fusion modules. In addition, we observe that the bilinear fusion plays a more important role than the feature selection since replacing the former causes more performance degradation.

\begin{table}[!t]
\renewcommand\arraystretch{1.05}
\setlength{\tabcolsep}{5.5pt}
\small
\centering
\caption{Bilinear fusion with different numbers of heads.}
\scalebox{1}{
\begin{tabular}{c|c|c|c|c}
\toprule
\#Heads ($k$) & Criteo & Avazu & MoiveLens & Frappe \\ \midrule
        1 & \texttt{OOM} & 0.7649 & 0.9691 & \textbf{0.9862} \\ 
        5 & 0.8141 & 0.7661 & 0.9707 & 0.9851 \\ 
        10 & 0.8144 & \textbf{0.7669} & \textbf{0.9724} & 0.9849 \\ 
        50 & \textbf{0.8148} & 0.7657 & 0.9703 & 0.9841  \\ \bottomrule   
\end{tabular}}
\label{tab:ablation_study_group} 
\vspace{-1ex}
\end{table}

\subsubsection{Effect of Multi-Head Bilinear Fusion}
We investigate the effect of our subspace grouping technique for bilinear fusion. 
Table~\ref{tab:ablation_study_group} shows the performances of FinalMLP by varying the number of subspaces (i.e, number of heads $k$) for bilinear fusion. \texttt{OOM} means that Out-Of-Memory error occurs in the setting. We found that using more parameters (i.e., smaller $k$) for fusion does not always lead to better performance. This is because an appropriate $k$ can help the model learn stream-level feature interactions from multiple views while reducing redundant interactions, similar to multi-head attention. One can achieve a good balance between effectiveness and efficiency by adjusting $k$ in practice.

\subsection{Industrial Evaluation}
We further evaluate FinalMLP in our production system for news recommendation, which serves millions of daily users. We first perform an offline evaluation using the training data from 3-day user click logs (with 1.2 billion samples). The AUC results are shown in Table \ref{tab:offline}. Compared to the deep BaseModel deployed online, FinalMLP obtains over one point improvement in AUC. We also make a comparison with EDCN~\cite{EDCN}, a recent work that enhances DCN~\cite{DCN} with interactions between two-stream networks. FinalMLP obtains additional 0.44 points improvement in AUC over EDCN. 
In addition, we test the end-to-end inference latency between receiving a user request and returning the prediction result. We can see that by applying our multi-head bilinear fusion, the latency can be reduced from 70ms (using 1 head) to 47ms (using 8 heads), achieving the same level of latency with the BaseModel (45ms) deployed online. Moreover, the AUC result also improves slightly by selecting an appropriate number of heads. We finally report the results of an online A/B test performed on July 18th$\sim$22nd, where the  results are shown in Table~\ref{tab:online_test}. FinalMLP achieves 1.6\% improvement in CTR on average, which measures the ratio of users' clicks over the total impressions of news. Such an improvement is significant in our production systems.


\begin{table}[!t]
    \small
    \renewcommand\arraystretch{1.15}
    \centering
    \caption{Offline results in production settings.}
    \begin{tabular}{c|c|c|c|c}
    \toprule
\multirow{2}{*}{} & \multirow{2}{*}{BaseModel} & \multirow{2}{*}{EDCN} & \multicolumn{2}{c}{FinalMLP} \\
                  &                            &                       & \multicolumn{1}{c}{\#Heads=1}      & \#Heads=8      \\ \midrule
        AUC & 71.78 & 72.22 & 72.83 & 72.93 \\ 
        $\Delta$AUC & -- & +0.44 & +1.05 & +1.15 \\\hline
        Latency & 45ms & -- & 70ms & 47ms \\\bottomrule
    \end{tabular}\label{tab:offline} 
\end{table}

\begin{table}[!t]
    \small
    \centering
    \caption{Online results of a five-day online A/B test.}
    \begin{tabular}{ccccccc}
    \toprule
         & Day1 & Day2 & Day3 & Day4 & Day5 & Average \\ \midrule
        $\Delta$CTR & 1.6\% & 0.6\% & 1.7\% & 1.5\% & 2.4\% & 1.6\% \\ \bottomrule
    \end{tabular}
    \label{tab:online_test} 
    \vspace{-1ex}
\end{table}


\section{Conclusion and Outlook}
In this paper, we makes the first effort to study a simple yet effective two-stream model, FinalMLP, that employs MLP in each stream for CTR prediction. To enhance the input differentiation of two streams and enable stream-level interaction, we propose stream-specific feature gating and multi-head bilinear fusion modules that are pluggable to improve the model performance. Our evaluation on four open datasets and in industrial settings demonstrates the strong effectiveness of FinalMLP. We emphasize that the surprising results of FinalMLP question the effectiveness and necessity of existing research in explicit feature interaction modeling, which should attract the attention of the community. We also envision that the simple yet effective FinalMLP model could serve as a new strong baseline for future developments of two-stream CTR models. Moreover, it is also an interesting future work to plug our feature gating and bilinear fusion modules into more two-stream CTR models.





\section*{Acknowledgments}
This work is supported by the Outstanding Innovative Talents Cultivation Funded Programs 2023 of Renmin Univertity of China. We gratefully acknowledge the support of MindSpore\footnote{\url{https://www.mindspore.cn}}, which is a new deep learning framework used for this research.

\bibliography{FinalMLP}
\balance
\end{document}